\documentclass[10pt,paperletter]{article}
\usepackage{graphicx}
\usepackage{amssymb}
\usepackage{amsmath}
\usepackage{latexsym}
\usepackage{appendix}

\title{\Large \textbf{Useful ratios between two-body nonleptonic and
semileptonic decays of $B$ mesons}}

\author{J. H. Mu\~noz$^{1,2}$\thanks{jhmunoz@ut.edu.co} $\;$ and N. Quintero$^{1,3}$\thanks{nquintero@fis.cinvestav.mx}\\
\small \textit{ $^1$Departamento de F\'isica, Universidad del Tolima, Apartado A\'ereo 546, Ibagu\'e, Colombia}\\
\small \textit{ $^2$ Centro Brasileiro de Pesquisas Fisicas, Rua Xavier Sigaud 150, 22290-180, Rio de Janeiro, RJ, Brazil}\\
\small \textit{ $^3$ Departamento de F\'isica, CINVESTAV, Apartado Postal 14-740, 07000 M\'exico D.F., M\'exico}}

\begin{document}
\maketitle

\begin{abstract}
We compute  important ratios between  decay widths  of some exclusive two-body nonleptonic and semileptonic $B$   decays, which could be   test of factorization hypothesis. We also present a summary of the  expressions of the  decay widths and differential decay rates of  these  decays, at tree level, including $l=0$ (ground state),  $l=1$ (orbitally excited)  and $n=2$ (radially excited)  mesons in the final state. From a general point of view,  we  consider  eight   transitions, namely   $H \to P, \; V, \; S, \; A, \;  A^{'}, \;  T, \;P(2S), \; V(2S)$.  Our analysis is carried out assuming  factorization hypothesis and using the WSB,   ISGW and  CLFA  quark models. 
\end{abstract}


\noindent
{\bf PACS numbers(s):} 13.20.-v, 13.25.-k, 12.39.-x

\vspace{0.3cm}

\section{Introduction}

Exclusive semileptonic and two-body nonleptonic  decays of heavy mesons offer a  good scenario  for studying, at theoretical and experimental levels,  CP violation and physics beyond the Standard Model.  Some of these channels provide methods for determining the angles of the unitarity triangle,  allow to study the role  of QCD and test some QCD-motivated models (see for example some recent reviews in  \cite{reviews}). These topics are of great interest in particle physics  and the knowledge of them will be improved with forthcoming experiments at Large Hadron Collider (LHC) \cite{LHC}.\\

The purpose of this paper is to compute useful ratios between two-body nonleptonic and semileptonic decays of heavy ($H$) mesons, at tree level,  that could be tested experimentally. Specifically, we work with  exclusive $B$  channels  although we  also consider a couple of $B_s$ processes. We assume naive factorization and use     the  WSB \cite{wsb},   ISGW \cite{isgw} and CLFA \cite{CLF} quark models.  \\

It is expected that  naive factorization approach  works reasonably well in decays where penguin and weak annihilation contributions are absent or suppressed, such as $B \to DK$  \cite {BABAR},   $K^{0} \to \pi\pi$,  $D^{0} \to K^{\pm}\pi^{\mp}$, $D^{0} \to K^{+}K^{-}, \; \pi^{+}\pi^{-}$ and $B_{s} \to J/ \psi \phi$ \cite{treeleveldominated}, $D^+ \to \overline{K}_0^{*0}\pi^+$ and $D_s^+ \to f_0\pi^+$ \cite{cheng-chiang} channels. Also, factorization assumption works well in  two-body hadronic decays of $B_{c}$ meson (without considering charmless modes) where the quark-gluon sea is suppressed in the heavy quarkonium \cite{Bc}.  \\

We also present an important summary and a general  analysis on the   expressions of the decay
widths and differential decay rates of two-body nonleptonic and semileptonic decays of heavy mesons, respectively, including $l=0, \; 1$ and $n=2$ mesons in the  final state. For $l=0$, we have considered pseudoscalar ($P$) and vector ($V$) mesons,  for $l=1$  we have included  orbitally excited ($p$-wave) scalar ($S$), axial-vector ($A,\;  A^{'}$) and tensor ($T$) mesons, and for $n=2$, we have studied  radially excited $P(2S)$ and $V(2S)$ mesons (see Table I).  We have classified   eight transitions, namely $H \to P,\; V, \; S, \; A, \; A^{'}, \; T$,  $P(2S), \; V(2S)$,   in three groups. It allows us to manipulate, in an easy way, all these decays. \\

The paper is organized  as follows:  In section 2 we present, in a general  way, the parametrization of  the hadronic matrix  element
$\langle M| J_{\mu}|H \rangle$ for eight cases. Section 3 contains
expressions for $\Gamma(H \rightarrow M_{1}M_{2})$ and $d\Gamma(H
\rightarrow Ml\nu)/dt$ and a brief discussion. In section 4, we analyze vector and axial contributions of the weak interaction to $H \rightarrow (P,V,S,A,A^{'},T)l\nu$ decays assuming a meson dominance model.  In section 5, we compute some important ratios between decay widths of exclusive $B$ (and $B_s$) decays, which allow us to get tests to factorization approach. Concluding remarks are presented in section 6. Finally, in the appendix we briefly mention  the quark models used in this work.

\begin{table}[ht]
{\small Table I. Classification of mesons considering the $n \; {}^{2s+1}L_{J}$ and  the $J^{PC}$ notations. $n$ is the radial quantum number, $l$ is the orbital angular momentum, $s$ is the spin, and $J$ is the total angular momentum. $P$ and $C$ are  parity and  charge conjugate operators, respectively. }
\par
\begin{center}
\renewcommand{\arraystretch}{1.5}
\begin{tabular}{|c|c|c|c|c|c|c|}
  \hline
 $n$ & $l$ & $s$ & $J$ & $n \; {}^{2s+1}L_{J}$ & $J^{PC}$ & Meson \\
  \hline\hline
  &$0$ & $0$ & $0$ & $1 \; {}^{1}S_{0}$ & $0^{-+}$ & Pseudoscalar $(P)$ \\
  \cline{3-7}
 &  & $1$ & $1$ & $1 \; {}^{3}S_{1}$ & $1^{--}$ & Vector $(V)$ \\
    \cline{2-7}
  &  & $0$ & $1$ & $1 \;  {}^{1}P_{1}$ & $1^{+-}$ & Axial-vector $(A')$ \\
    \cline{3-7}
 $1$ & $1$ &   & $0$ & $1 \; {}^{3}P_{0}$ & $0^{++}$ & Scalar $(S)$ \\
  \cline{4-7}
  &  & $1$ & $1$ & $1 \; {}^{3}P_{1}$ & $1^{++}$ & Axial-vector $(A)$ \\
   \cline{4-7}
  &  &   & $2$ & $1 \; {}^{3}P_{2}$ & $2^{++}$ & Tensor $(T)$ \\
  \hline
  $2$ & $0$ & $0$ & $0$ & $2 \; {}^{1}S_{0}$ & $0^{-+}$ & $P(2S)$ \\
  \cline {3-7}
  & & $1$ & $1$ & $2 \; {}^{3}S_{1}$ & $1^{--}$ & $V(2S)$\\
  \hline
\end{tabular}
\end{center}
\end{table}


\section{Hadronic Matrix Elements}

In this section, we present  the parametrizations of the eight  $H \to M$ transitions, where $H$ denotes a pseudoscalar heavy meson and    $M$ can be a $P$, $V$, $S$, $A$, $A^{'}$, $T$, $P(2S)$, $V(2S)$ meson,  classified in three groups\footnote{We use the ISGW model \cite{isgw} because it provides all the parametrizations considered in this work.}. In the first case, the $M$ meson   has $J=0$, in the second,  $J=1$, and in the third group  $J=2$.


\subsection{$H \to M(J=0)$ transition}

In this group, there are three transitions if  $M$ is a meson with $J=0$ (see Table I): $M$ can be the pseudoscalar $P$ meson, or  the  scalar $S$ meson, which is an  orbitally excited meson, or  the radially excited meson $P(2S)$. The hadronic matrix element $\langle M|J_{\mu}|H\rangle$ for $M = P, \; S, \; P(2S)$ has the same   Lorentz structure and it is  defined  as follows \cite{isgw}:

\begin{eqnarray}\label{BP}
\langle M(p_M)|J_{\mu}| H(p_H)\rangle &\equiv& F_{+} \; (p_H+p_M)_\mu \;  + \; F_{-} \; (p_H-p_M)_\mu,
\end{eqnarray}

where $J_{\mu}$ is the $V_{\mu} - A_{\mu}$ weak current,  $p_{H(M)}$ is the 4-momentum of the meson $H(M)$, $F_{+} $ and $F_{-}$  are  form factors.  Following the notation displayed in  appendix of the ISGW model \cite{isgw}, these form factors are:

\begin{itemize}
\item For $M = P$: $\langle P|J_{\mu}|H\rangle \equiv \langle P|V_{\mu}|H\rangle$, $F_{+} = f_{+}$ and $F_{-} = f_{-}$.
\item For $M = S$: $\langle S|J_{\mu}|H\rangle \equiv -\langle S|A_{\mu}|H\rangle$, $F_{+} = u_{+}$ and $F_{-} = u_{-}$.
    \item For $M = P(2S)$: $\langle P(2S)|J_{\mu}|H\rangle \equiv \langle P(2S)|V_{\mu}|H\rangle$, $F_{+} = f^{'}_{+}$ and $F_{-} = f^{'}_{-}$.
\end{itemize}

It is important to note that the parity operator requires that  $\langle P|A_{\mu}|H\rangle = 0$ and $\langle S|V_{\mu}|H\rangle = 0$.\\

Ref. \cite{wsb} uses a different parametrization for $\langle P|J_{\mu}|H\rangle$ using dimensionless $F_{1}$ and $F_{0}$ form factors. It is possible to transform $(F_{1},\; F_{0}) \; \to \; (f_{+}, \; f_{-})$ using the relations showed in  the appendix.


\subsection{$H \to M(J=1)$ transition}

Considering the $M$ meson  with $J=1$, this group  has four transitions (see Table I):  $M = V, \; A, \; A^{'}, \; V(2S)$. The hadronic matrix element $\langle V\; (A(1 \; ^{3}P_{1}), A(1 \; ^{1}P_{1}), V(2S))\; |J_{\mu}|H\rangle$ can be parametrized by means of the following linear combination which is Lorentz-covariant \cite{isgw}:
\begin{eqnarray}\label{BV-ISGW} \nonumber
\langle M(p_M,\epsilon)|J_\mu| H(p_H)\rangle &\equiv& iG\varepsilon_{\mu \nu\rho\sigma}\epsilon^{*\nu}(p_H + p_M)^{\rho}(p_H - p_M)^{\sigma} + F\epsilon^{*}_{\mu} \\
 && + \; A_+ (\epsilon^{*}.p_H)(p_H + p_M)_{\mu} + A_- (\epsilon^{*}.p_H)(p_H - p_M)_{\mu},
\end{eqnarray}
where $G$, $F$, and $A_{\pm}$ are form factors, $\epsilon$ is the polarization vector of  meson $M$ and $p_{H(M)}$ is the 4-momentum of the meson $H(M)$.  Following the notation used in  the appendix of the ISGW model \cite{isgw}, these form factors are:

\begin{itemize}
\item For $M = V$:  $G = g$,  $F = -f$, $A_+ = -a_+$ and $A_- = -a_-$.
\item For $M = A(1 \; ^{3}P_{1}) \equiv A$: $G = -q$,  $F = l$, $A_+ = c_+$ and $A_- = c_-$.
    \item For $M = A(1 \; ^{1}P_{1}) \equiv A^{'}$:  $G = -v$,  $F = r$, $A_+ = s_+$ and $A_- = s_-$.
    \item For $M = V(2S)$: $G = g^{'}$,  $F = -f^{'}$, $A_+ = -a_+^{'}$ and $A_- = -a_-^{'}$.
\end{itemize}

The parametrization of  the matrix element for the $H \to A$ transition  has the same structure that  the matrix element of the $H \to V$ transition just interchanging the role of vector and axial currents: $ \langle V|V_\mu (A_\mu)|H\rangle \leftrightarrow \langle A|A_\mu (V_\mu)|H\rangle$.\\

Ref. \cite{wsb} (\cite{CLF}) works with  another parametrization for the $H \to V \; (A)$ transition, which is very useful  because it allows to write the decay width of two-body nonleptonic decays of heavy mesons as a  function of helicity form factors (see for example the Refs. \cite{wsb, ebert2007}). It is easy to transform  the parametrization given by the equation (\ref{BV-ISGW}) into the parametrization given in the Refs. \cite{wsb, CLF} by using  the relations between form factors showed in the appendix.


\subsection{$H \to M(J=2)$ transition}

This group contains only one transition (see Table I): when $M$ is a tensor meson ($T$), which is a $p$-wave.  The Lorentz-covariant parametrization of the  hadronic matrix element $\langle T|J^{\mu}|H \rangle$  given in the ISGW model is \cite{isgw}:

\begin{eqnarray}
\langle T(p_{T},\epsilon)|J^{\mu}|H(p_H)\rangle = i h(q^{2}) \varepsilon^{\mu\nu\rho\sigma} \epsilon_{\nu\alpha}p_{H}^{\alpha} (p_{H}+p_{T})_{\rho}(p_{H}-p_{T})_{\sigma} \nonumber\\
  - \ \   k(q^{2}) \epsilon^{*\mu\nu} (p_{H})_{\nu} \ \ + \ \ \epsilon_{\alpha\beta}^{*}p_{H}^{\alpha}p_{H}^{\beta}  \left[ b_{+}(q^{2})(p_{H}+p_{T})^{\mu} + b_{-}(q^{2})(p_{H}-p_{T})^{\mu} \right],
\end{eqnarray}

\noindent where $\epsilon_{\nu\alpha}$ is the polarization tensor of the tensor meson, $p_{H(T)}$ is the momentum of the heavy meson $H(T)$, and  $h, k, b_{\pm}$ are  form factors.   $k$ is dimensionless and $h, b_{\pm}$ have dimensions of GeV$^{-2}$.\\

In the literature \cite{BT1, BT2}, there is  another parametrization of $\langle T|J^{\mu}|H \rangle$, which is constructed in analogy with  the parametrization of $\langle V|J^{\mu}|H \rangle$ given in  Ref. \cite{wsb},   using the tensor polarization $\epsilon_{\mu\nu}$ of the $T$ meson.


\section{ $d\Gamma(H
\rightarrow Ml\nu)/dt$ and $\Gamma(H \rightarrow M_{1}M_{2})$}

In this section we collect, in a compact  form, using the classification of  the last section, the expressions, at tree level, of the differential decay rate of $H \rightarrow Ml\nu_{l}$
(see Table II) and  the decay width of $H \to MM^{'}$ (see Table III), where $H$ is a heavy meson ($D$, $D_{s}$, $B$, $B_{s}$ or $B_{c}$), and $M \; (M^{'})$ can be any of  the eight mesons  $P, \; V, \; S$, $A, \; A^{'}, \; T$, $P(2S), \; V(2S)$.\\

In  the first row of  Table II, we display  the differential decay rate of the semileptonic $H \rightarrow Ml\nu_{l}$ decay, where  $M$ is a meson with $J=0$, i.e, $M = P, \; S,\; P(2S)$, using the parametrization given in the WSB model \cite{wsb}. The second row shows  the differential decay rate of $H \rightarrow M l\nu_{l}$, where  $M$ is a meson with $J=1$, i.e, $M = V, \; A, \; A^{'}, \; V(2S)$, using  parametrizations given in  the WSB \cite{wsb} and  ISGW \cite{isgw} quark models, and in  the last row we give the  differential decay rate for $H \rightarrow T(J=2) l\nu_{l}$ using the parametrization of the  ISGW model \cite{isgw}.

\begin{table}[ht]
{\small Table II. Differential decay widths of $H \rightarrow
(P, \; V, \; S, \; A, \; A^{'}, \; T, \; P(2S), \; V(2S))l\nu_{l}$.}
\par
\begin{center}
\renewcommand{\arraystretch}{2}
\begin{tabular}{||c|c||}
  \hline
 $H \rightarrow Ml\nu_{l}$  & $d\Gamma(H \rightarrow Ml\nu_{l})/dt$ \\
  \hline\hline
  $H \rightarrow (P, \; S, \; P(2S))l\nu_{l}$ & $\zeta \left[A(t)|F_{1}^{HM}(t)|^{2}\lambda^{3/2} + B(t) |F_{0}^{HM}(t)|^{2}\lambda^{1/2} \right]$ \\
  \hline
   & $\zeta\mathcal{G}(t)$ \\
  \cline{2-2}
  $H \rightarrow (V, \; A, \; A^{'}, \; V(2S))l\nu_{l}$ & $\zeta t \lambda^{1/2} \left[ |H_{+}(t)|^{2} + \left|H_{-}(t)\right|^{2}+ |H_{0}(t)|^{2}\right]$ \\
  \cline{2-2}
   & $\zeta \left\{ \varphi(t)\lambda^{5/2} + \rho(t)\lambda^{3/2}+ \theta(t)\lambda^{1/2} \right\}$ \\
  \hline
  $H \rightarrow Tl\nu_{l}$ & $\zeta \left\{ \alpha(t)\lambda^{7/2} + \beta(t)\lambda^{5/2}+ \gamma(t)\lambda^{3/2} \right\}$ \\
  \hline
\end{tabular}
\end{center}
\end{table}

In Table II, $\lambda =
\lambda(m_{H}^{2},m_{M}^{2},t)$, where $\lambda=\lambda(x,y,z)=x^{2}+y^{2}+z^{2}-2xy-2xz-2yz$ is the
triangular function, $t = (p_H -p_M)^{2}$ is the momentum transfer and $H_{\pm,0}$ are helicity form factors \cite{wsb}. The  factor $\zeta$ and  functions $A(t)$, $B(t)$, $\mathcal{G}(t)$, $\varphi(t)$, $\rho(t)$, $\theta(t)$, $\alpha(t)$, $\beta(t)$ and $\gamma(t)$  are defined by:  
\begin{equation}
\zeta = \frac{G_{F}^{2} |V_{q^{'}q}|^{2}}{192\pi^{3}m_{H}^{3}},
\end{equation}

\begin{equation}
A(t) = \left( \frac{t-m_{l}^{2}}{t} \right)^{2}
\left(\frac{2t+m_{l}^{2}}{2t} \right),
\end{equation}

\begin{equation}
B(t) = \frac{3}{2} m_{l}^{2} \left( \frac{t-m_{l}^{2}}{t}
\right)^{2} \frac{(m_{H}^{2}- m_{P}^{2})^{2}}{t},
\end{equation}

\begin{align}
\mathcal{G}(t) = \; & \left[
\frac{2t|V(t)|^{2}}{(m_{H}+m_{V})^{2}} +
\frac{(m_{H}+m_{V})^{2} |A_{1}(t)|^{2}}{4m_{V}^{2}}
- \frac{(m_{H}^{2}-m_{V}^{2}-t)
A_{1}(t)A_{2}(t)}{2m_{V}^{2}} \right] \lambda^{3/2}
\nonumber \\
& \;+ \frac{|A_{2}(t)|^{2}}{4m_{V}^{2}
(m_{H}+m_{V})^{2}} \lambda^{5/2} +
3t (m_{H}+m_{V})^{2} |A_{1}(t)|^{2} \lambda^{1/2},
\end{align}

\begin{equation}\label{1}
\varphi(t) = \frac{s_{+}^{2}}{4m_{A}^{2}},
\end{equation}

\begin{equation}\label{2}
\rho(t) = \frac{1}{4m_{A}^{2}}
\left[r^{2}+8m_{A}^{2}tv^{2}+2(m_{H}^{2}-m_{A}^{2}-t)rs_{+}\right],
\end{equation}

\begin{equation}\label{3}
\theta(t) = 3 t\;r^{2},
\end{equation}

\begin{equation}
\alpha(t) = \frac{b_{+}^{2}}{24m_{T}^{4}},
\end{equation}

\begin{equation}
\beta(t) = \frac{1}{24m_{T}^{4}} \left[k^{2}+6m_{T}^{2}th^{2}+2(m_{H}^{2}-m_{T}^{2}-t)kb_{+}\right],
\end{equation}

\begin{equation}
\gamma(t) = \frac{5 tk^{2}}{12m_{T}^{2}},
\end{equation}

where $G_F$ is the Fermi constant, $m_{H(P, \; V, \; A, \; T)}$ is the   mass of the $H(P, \; V, \; A, \; T)$ meson, $m_l$ is the mass of the lepton,   $V(t)$ and $A_{1,2}(t)$ are  form factors \cite{wsb}, $\varphi(t)$, $\rho(t)$ and $\theta(t)$ are quadratic
functions of the  form factors $s_{+}$, $r$ and $v$ ($c_{+}$, $l$ and $q$) for $H \rightarrow A(^{1}P_{1})l\nu$ ($H \rightarrow A(^{3}P_{1})l\nu$), $\alpha(t)$, $\beta(t)$
and $\gamma(t)$ are quadratic functions \cite{alexander} of  the form factors $k$,
$b_{+}$ and $h$. All these form factors are explicitly given  in the appendix B of the Ref. \cite{isgw}.\\

The dependence of $d\Gamma(H \rightarrow Ml\nu)/dt$ with   $\lambda$ $(|\overrightarrow{p}|=\lambda^{1/2}/2m_H$, where $\overrightarrow{p}$ is the three-momentum of the $M$ meson  in the $H$ meson rest frame) is given by,
\begin{equation}\nonumber
d\Gamma/dt \sim \lambda^{l+\frac{1}{2}},
\end{equation}
where $l$ is the orbital angular momentum of the wave at which  the  particles in  the final state can be coupled.  Assuming conservation of total angular momentum $J$ and  a meson dominance model we  can find   specific values for $l$ in each exclusive  $H \rightarrow Ml\nu$ decay. Thus, in $H \rightarrow M(J=0)l\nu$  the particles in the final state are coupled to  $l=0, 1$ waves (see the first row in Table II). When $m_l \approx 0$ ($l = e, \mu$), the coefficient $B(t)$ vanishes, so the contribution of the $s$-wave is negligible; in $H \rightarrow M(J=1)l\nu$  the particles in the final state can be coupled to  $l=0, 1, 2$ waves (see the second row in Table II); and in $H \rightarrow T(J=2)l\nu$ to  $l=1,2,3$ waves (see the last row in Table II).  \\

It is also possible to write in a compact expression the differential decay rate of the  semileptonic  $H \rightarrow Ml\nu_{l}$ decay, where $M$ is a p-wave (orbitally excited) meson: scalar, vector-axial or tensor meson, in terms of helicity amplitudes (see  Ref. \cite{ebert2010}).  \\


As for  two-body nonleptonic decays of heavy mesons,   the effective weak Hamiltonian $\mathcal{H}_{eff}$ has contributions from  current-current (tree), QCD penguin and electroweak penguin operators \cite{buras}. In general, $\mathcal{H}_{eff} \approx \sum_{i} C_i(\mu)\mathcal{O}_i$, where $C_i(\mu)$ are the Wilson coefficients and $\mathcal{O}_i$ are  local operators. The amplitude for the $H \to M_1M_2$ decay is
\begin{equation}
\mathcal{M}(H \to M_1M_2)  \approx \sum_{i} C_i(\mu) \langle  \mathcal{O} \rangle_i.
\end{equation}

In the scenario of naive factorization, it is assumed that
\begin{equation}
\mathcal{M}(H \to M_1M_2) \approx C_i(\mu) \langle M_2 |(J_{1i})_{\mu} | 0 \rangle  \langle M_1 |(J_{2i})^{\mu} | H \rangle + (M_1 \leftrightarrow M_2),
\end{equation}
where $J_{\mu}$ is the  $V_{\mu}-A_{\mu}$ weak current and the hadronic matrix element of a four-quark operator is written as the  product of a decay constant and form factors \cite{factorization}.\\

This factorization presents a difficulty because the Wilson coefficients are $\mu$ scale and renormalization scheme dependent while $\langle  \mathcal{O} \rangle_i$ are $\mu$ scale and renormalization scheme independent,  so clearly the physical amplitude depends on the   $\mu$ scale. The naive factorization disentangles the short-distance effects from the long-distance sector assuming that $\langle  \mathcal{O} \rangle_i$, at  $\mu$ scale,  contain nonfactorizable  contributions in order to cancel the $\mu$ dependence and the scheme dependence of $C_i(\mu)$. Thus, the naive factorization is an approximation because it does not consider possible QCD interactions between the meson $M_2$ and the $H$ and $M_1$ mesons. In general,  it does not work  in all two-body heavy meson decays \cite{factorization}.  \\

Assuming  naive factorization, we have considered only those decays which are produced by the color-allowed external  $W$-emission tree diagram or the color-suppressed internal  $W$-emission diagram. It is expected that  naive factorization works reasonably well in decays where penguin and weak annihilation contributions are absent or negligible, as for example in $B \to DK$  \cite {BABAR},   $K^{0} \to \pi\pi$,  $D^{0} \to K^{\pm}\pi^{\mp}, \;   K^{+}K^{-}, \; \pi^{+}\pi^{-}$ and $B_{s} \to J/ \psi \phi$ \cite{treeleveldominated}, $D^+ \to \overline{K}_0^{*0}\pi^+$ and $D_s^+ \to f_0\pi^+$ \cite{cheng-chiang} channels. Also, factorization assumption works well in  two-body hadronic decays of $B_{c}$ meson (except in charmless  processes, because they  are produced only by annihilation contributions) where the quark-gluon sea is suppressed in the heavy quarkonium \cite{Bc}. We have used
the notation $H \rightarrow M_{1},  M_{2}$ \cite{chen} to mean that
$M_{2}$  is factorized out under   factorization approximation,  i.e., $M_2$ arises from  the vacuum. For $H \to TM$ decays there is not any possibility to produce the $T$ meson from the vacuum with the $V-A$ current, because $\langle  T|(V-A)_{\mu}|0   \rangle \equiv 0$. So,  this decay has only the contribution $H \to T, M$. Recently, it has been reported that it is possible to produce tensor mesons from the vacuum  involving covariant derivatives \cite{BT2,T-vacio}.\\

Using the  parametrizations given in section 2 for eight transitions, namely $H \to M(J=0, \; 1, \; 2)$, we display, in  Table III, expressions of   decay widths for 40 different types of  $H(q_H\bar{q}') \rightarrow
M_{1}(q\bar{q}')M_{2}(q_{i}\bar{q}_{j})$ decays, which are produced by the $q_H \to q\overline{q}_jq_i$ transition. \\

In  the first row of  Table III, we show the decay width for six different types of channels: $H \to P,P^{'}; \; P,P^{'}(2S); \; S,P^{'}; \; S,P^{'}(2S); \; P(2S),P^{'}; \; P(2S),P^{'}(2S)$. They are produced by the $H \to M(J=0)$ transition.   The hadronic matrix elements $\langle P (S,\;  P(2S)) | J_{\mu}|H \rangle$, which are neccesary in order to calculate the decay width, have the same parametrization. In this case, we have used the parametrization presented in \cite{wsb}. In these decays the particles in the final state are coupled to a $s$- wave because $\Gamma \sim \lambda^{0+1/2}$. In a similar way, in  the second row of Table III, we display the decay width of nine different modes: $H \to P,V; \; P,A; \; P, V(2S); \; S,V; \; S,A; \; S, V(2S); \; P(2S),V; \; P(2S),A$;  $P(2S), V(2S)$. These nine channels have in common the $H \to M(J=0)$  transition. In these decays, the particles in the final state are coupled to a $p$-wave ($l=1$).\\

In  the third row of  Table III, we present the decay width for eight different types of decays: $H \to V,P; \; V, P(2S); \; A,P; \; A, P(2S); \; A^{'},P; \; A^{'}, P(2S); \; V(2S),P; \; V(2S), P(2S)$. The  hadronic matrix elements $\langle V (A, \; A^{'}; V(2S))| J_{\mu}|H \rangle$,  which correspond to the  $H \to M(J=1)$ transition, have a similar parametrization.  The particles in the final state in these decays are coupled to a $p$-wave ($l=1$). In  the fourth row of  Table III, we display the decay width for twelve different decays: $H \to V_1,V_2; \;  V_1, A_2; \; V_1, V_2(2S); \; A_1,V_2$;    $A_1, A_2; \; A_1, V_2(2S); \; A_1^{'},V_2; \;  A_1^{'}, A_2; \; A_1^{'}, V_2(2S)$;  $V_1(2S),V_2$;   $V_1(2S), A_2; \; V_1(2S), V_2(2S)$. They also arise from  the $H \to M(J=1)$ transition\footnote{For the $H \to A, \;  A'$ transitions it is required to interchange the role of vector and axial currents in order to obtain the specific expressions displayed in Tables II and III.}. The two $J=1$ particles in the final state can be coupled to $l=0, 1, 2$ waves.\\

In the fifth row of  Table III, we show the decay widht for the $H \to T, P( P(2S))$ channels, which are produced by the  $H \to T$ transition.  We have used the parametrization for $\langle T| J_{\mu}| H \rangle$ given in the  Ref. \cite{isgw}.  In this case, the particles in the final state can be coupled to a $l=2$ wave. Using the same parametrization, we present in the last row of  Table III,  the decay width for three different modes:  $H \to T, V (A, \; V(2S))$. In this case, the particles in the final state can be coupled to $l=1, 2, 3$ waves.\\

In Table III, all
 form factors and the function $\lambda$ are evaluated in $m_{M_2}^{2}$ because  the momentum transfer $t = (p_H - p_1)^{2}=p_2^{2}=m_{M_2}^{2}$.   
$\xi^{(M_{2})}$,  $\mathcal{F}^{H \rightarrow T}$ and  the decay constants are given by
\begin{eqnarray}
\xi^{(M_{2})} &=&
\frac{G_{F}^{2}|V_{qq_H}|^{2}|V_{q_{i}q_{j}}|^{2} a_{1(2)}^{2}
f_{M_{2}}^{2}}{32\pi m_{H}^{3}}, \\
\mathcal{F}^{H \rightarrow T}(m_{P}^{2}) &=& k + (m_{H}^{2}-
m_{T}^{2}) b_{+} + m_{P}^{2}b_{-}, \\
\langle M(p) | J_{\mu} | 0   \rangle &=& if_Mp_{\mu}, \ \ \ \    M = P, \;  P(2S),\\
 \langle M(p,\epsilon) | J_{\mu} | 0 \rangle &=& f_Mm_M\epsilon_{\mu}, \ \ \ \    M = V, \; A,\; V(2S),
\end{eqnarray}
where $k$ and $b_{\pm}$  are form factors given in  the ISGW model
\cite{isgw},
evaluated at $t = m_{P}^{2}$,  $a_{1(2)}$ are the QCD factors,  and $|V_{qq_H}|$ and $|V_{q_iq_j}|$ are the appropriate CKM factors.\\

 Finally,  we do not consider decays where   a tensor meson, or a scalar meson or an axial-vector meson $1 \; ^{1}P_1$ arises from the vacuum.  In the first case, as we mentioned before, $\langle T|J_{\mu}  |0\rangle \equiv 0$;  in the second case, the decay constant of the scalar mesons, defined as $\langle S|J_{\mu}|0\rangle = f_Sp_{\mu}$ vanishes or is small (of the order of $m_d - m_u$, $m_s - m_{u,d}$); and  in the last case, the decay constant of the $1 \; ^{1}P_1$ meson vanishes in the SU(3) limit by $G$- parity \cite{cheng2009}. 

\begin{table}[ht]
{\small Table III. Decay widths of $H \rightarrow M_1M_2$, where $M_{1,2} = P, \; V, \; S, \; A, \; A^{'}, \; T, \; P(2S), \; V(2S)$}
\par
\begin{center}
\renewcommand{\arraystretch}{1.5}
\begin{tabular}{||c|c||}
  \hline
  $H \rightarrow M_{1},M_{2}$ & $\Gamma(H \rightarrow M_{1},M_{2})$ \\
  \hline\hline
  $H \rightarrow (P_{1}, S_1, P_1(2S)), \; (P_{2}, P_2(2S))$   & $\xi^{(M_{2})}(m_{H}^{2}-m_{M_1}^{2})^{2} |F_{0}^{HM_1}(m^{2}_{M_2})|^{2} \lambda^{1/2}$ \\
  \hline
  $H \rightarrow (P_{1}, S_1, P_1(2S)), \; (V, A, V(2S))$ & $\xi^{(M_2)}|F_{1}^{HM_1}(m^{2}_{M_2})|^{2} \lambda^{3/2}$ \\
  \hline
  $H \rightarrow (V, A, A^{'},V(2S)), \; (P, P(2S))$ & $\xi^{(M_2)}|A_{0}^{HM_1}(m^{2}_{M_2})|^{2} \lambda^{3/2}$ \\
  \hline
  $H \rightarrow (V_1, A_1, A_1^{'},V_1(2S)), \; (V_{2}, A_2, V_2(2S))$ & $\xi^{(M_{2})} \mathcal{G}(t=m_{V_2}^{2})$ \\
  \cline{2-2}
   & $\xi^{(M_{2})} m_{2}^{2} \lambda^{1/2} \left[|H_{+}(m^{2}_{M_2})|^{2} + |H_{-}(m^{2}_{M_2})|^{2} + |H_{0}(m^{2}_{M_2})|^{2} \right]$ \\
  \hline
  $H \rightarrow T, \; (P, P(2S))$ & $\xi^{(M_2)} (1/24m_{T}^{4}) |\mathcal{F}^{H \rightarrow T}(m^{2}_{M_2})|^{2} \lambda^{5/2}$ \\
  \hline
  $H \rightarrow T,\; (V, A, V(2S))$ & $\xi^{(M_2)} \left[ \alpha(m^{2}_{M_2})\lambda^{7/2} + \beta(m^{2}_{M_2})\lambda^{5/2}+ \gamma(m^{2}_{M_2})\lambda^{3/2} \right]$ \\
  \hline
  \end{tabular}
\end{center}
\end{table}


\section{Contributions of the vector and axial couplings}

In this section,  we illustrate how  the particles in  the final state of $H \rightarrow Ml\nu$ and $H \rightarrow M_1M_2$ decays can be coupled to specific waves,  obtain  the quantum numbers of  the poles that appear in  the monopolar form factors, and explain the correspondence between the form factors and the respective waves in the final state.    We show that these  numbers depend on    the vector and axial couplings of the  weak interaction. Let us  consider the decay chain $H \rightarrow MM^{*} \rightarrow MW^{*} \rightarrow Ml\nu(MM')$, where  $W^{*}$ is  the off-shell intermediate  boson of the weak interaction. We need to  combine parity and total angular momentum conservations  in  the strong $H \rightarrow MM^{*}$  process.   \\

In Table IV, we show the specific waves in which particles in the final state of  $H \rightarrow (P,V,S,A,T)l\nu$ decays can be coupled  and determine if they come from the vector or axial contributions. We must  keep in mind that the off-shell $W^*$ boson  has spin $0$ or $1$. Thus, in the vector coupling there are two possibilities: $S_{W^*}=0$ with $P_{W^*}=+1$, and $S_{W^*}=1$ with $P_{W^*}=-1$ ($S_{W^*}$ and $P_{W^*}$ denote spin and parity of  $W^*$,  respectively). In a similar way, in  the axial coupling there are two options: $S_{W^*}=0$ with $P_{W^*}=-1$ and $S_{W^*}=1$ with $P_{W^*}=+1$. Thus, there are four cases for the $W^{*}$ boson: $J^P = 0^{+}$, $1^{-}$, $0^{-}$ and $1^{+}$. They are displayed in the second column of Table IV. Assuming   total angular momentum  and parity conservations  of the strong  $H \rightarrow M M^{*}$ process, we obtain the values of the orbital angular momentum $l$ of the particles in the final state of $H \to M l\nu$ (see Table IV). These values can be verified with the exponent $l+ \frac{1}{2}$ of $\lambda$ in  the expressions for $\frac{d\Gamma}{dt}$ in Table II. We can see in the third (fourth) and the fifth (sixth) columns in Table IV, that the vector and axial contributions interchange their roles in $H \to Pl\nu$ ($H \to Vl\nu$) and  $H \to Sl\nu$  ($H \to Al\nu$), respectively.  
\begin{table}[ht]
{\small Table IV. Vector and axial contributions to semileptonic $H \rightarrow (P,V,S,A(^{3}P_{1} \text{or} ^{1}P_1),T)l\nu$ decays.}
\begin{center}
\renewcommand{\arraystretch}{1.3}
\begin{tabular}{||c|c|c|c|c|c|c||}
  \hline
  Contribution & $J^{P}$ of $W^{\ast}$ & $H \rightarrow Pl\nu$ & $H \rightarrow Vl\nu$ & $H \rightarrow Sl\nu$ & $H \rightarrow Al\nu$ & $H \rightarrow Tl\nu$ \\
  \hline\hline
  Vector & $0^{+}$ & $l=0$ &  &  & $l=1$ & \\
  \cline{2-7}
   & $1^{-}$ & $l=1$ &$l=1$  &  & $l=0$, $l=2$&$l=2$\\
  \hline\hline
  Axial & $0^{-}$ &  & $l=1$ &  $l=0$&  & $l=2$\\
  \cline{2-7}
   & $1^{+}$ &  & $l=0$, $l=2$ &$l=1$  & $l=1$  &$l=1$, $l=3$\\
  \hline\hline
\end{tabular}
\end{center}
\end{table}

In Table V, we show the respective form factors  with the corresponding poles in $H \rightarrow P(V)l\nu$  decays. In the second column, we list the quantum numbers $J^{P}$ of  poles, which are the same $J^{P}$ options for the off-shell $W^*$ boson (see the second column in Table IV). In this case, we must check  the form factors that appear in  the parametrization of the hadronic matrix elements $\langle M| V_{\mu}|H\rangle$ and $\langle M| A_{\mu}|H\rangle$  for $M = P,V$. Following this idea, we obtain the  quantum numbers of  the poles for $H \rightarrow Ml\nu$ where $M$ is a $p$-wave meson:
for $H \rightarrow Sl\nu$ the poles are $0^{-}$ and $1^{+}$;
 for  $H \rightarrow Al\nu$,  the poles are $0^{+}$, $1^{-}$ and $1^{+}$;
and  for $B \rightarrow Tl\nu$ the poles are $1^{-}$, $0^{-}$ and  $1^{+}$.
These values are important if we are interested in constructing a  quark model with monopolar form factors for  $H \rightarrow S, \; A, \; T$  transitions.\\

 Let us illustrate, as an example, the situation on $H \rightarrow Pl\nu$  from Tables IV and V. This decay has two contributions: $l=0$ and $l=1$ (see  exponents of $\lambda$ in   Table II) which arise from  the vector coupling of the weak current (see  Table IV). The respective poles have quantum numbers $0^{+}$ and $1^{-}$  and  the form factors are $F_{0}$ and $F_{1}$ (see Table V).

\begin{table}[ht]
{\small Table V. Form factors and the  vector and axial contributions of the weak interaction to  $H \rightarrow (P,V)l\nu$ decays.}
\par
\begin{center}
\renewcommand{\arraystretch}{1.3}
\begin{tabular}{||c|c|c|c||}
  \hline
  Contribution & $J^{P}$ of Pole & $H \rightarrow Pl\nu$ & $H \rightarrow Vl\nu$  \\
  \hline\hline
  Vector & $0^{+}$ & $F_0(t)$&   \\
  \cline{2-4}
   & $1^{-}$ & $F_1(t)$ &$V(t)$\\
  \hline\hline
  Axial & $0^{-}$ &  &$A_0(t)$ \\
  \cline{2-4}
   & $1^{+}$ &  &$A_1(t)$, $A_2(t)$, $A_3(t)$ \\
  \hline\hline
\end{tabular}
\end{center}
\end{table}


\section{Useful ratios}

In this section, we present some ratios between  exclusive semileptonic and two-body nonleptonic decays of
$B$ and $B_s$ mesons, using the expressions for $d\Gamma(H \rightarrow Ml\nu)/dt$ and $\Gamma(H \rightarrow M_1,M_2)$ (see Tables II and III, respectively),  that could be a test of factorization hypothesis with  forthcoming measurements at LHC. We have worked with decays  where it is expected that naive factorization works well. In
order to obtain the numerical values presented in this section, we have evaluated the form factors in the WSB \cite{wsb} and CLFA \cite{CLF} quark models  and taken from the Particle Data Group \cite{pdg} the values of the CKM factors, branching ratios,  masses and  mean lifetime of mesons. \\

\noindent
\textbf{5.1} \textit{$B \to P, M(c\overline{c})$  decays:} Let us consider exclusive   two-body nonleptonic $B$ decays with  orbitally  or  radially excited  charmonium mesons in the final state, which are produced  by the color suppressed $b \to c\overline{c}s(d)$ transition.  The   following ratio
\begin{equation}\nonumber
\frac{\Gamma(B^+ \to P^+,M_1(c\overline{c}))}{\Gamma(B^+ \to P^+,M_2(c\overline{c}))} = (\text{kinematical factor}) \left (  \dfrac{f_{M_1}}{f_{M_2}} \right)^2  \left|  \frac{F_{0(1)}^{B \to P}(m^2_{M_1})}{F_{0(1)}^{B \to P}(m^2_{M_2})}     \right|^2,
\end{equation}
allows  to  obtain  decay constants of  charmonium mesons. The form factor $F_{0(1)}$ corresponds when  $M_1$ and $M_2$  are $J=0(1)$ mesons. \\

Evaluating the form factors in the  CLFA model  \cite{CLF}, we obtain
\begin{eqnarray}\nonumber
 \frac{f_{J/\psi}}{f_{\psi(2S)}} &=& 1.15 \pm 0.07 \; (1.29 \pm 0.17), \nonumber \\
\frac{f_{J/\psi}}{f_{\chi_{c1}(1P)}} &=& 1.41 \pm 0.13 \; (1.51 \pm 0.32), \nonumber \\
 \frac{f_{\eta_c}}{f_{\eta_c(2S)}} &=& 1.65 \pm 1.27, \nonumber \\
\frac{f_{\eta_c}}{f_{\chi_{c0}(1P)}} &=& 2.63 \pm 0.52, \nonumber
\end{eqnarray}
  taking $(M_1 = J/\psi,  \; M_2 = \psi(2S),  \;  P = K (\pi))$, $(M_1 = J/\psi, \;  M_2 = \chi_{c1}(1P), \; P = K  (\pi))$,  $(M_1 = \eta_c, \;  M_2 = \eta_c(2S),  \;   P = K)$, and
    $(M_1 = \eta_c, \; M_2 = \chi_{c0}(1P), \; P = K)$,   respectively. The most important sources of uncertainties come from experimental values of  branching ratios and  form factors. However, the error in the last ratios is dominated by the uncertainty in the  branching ratios. These quotients between decay constants with orbitally and radially excited charmonium states are a good test of the factorization hypothesis.\\

On the other hand, taking $f_{J/\psi} = (416.3 \pm 5.3)$ MeV \cite{Colangelo-Fazio-Wang, Wang-Shen-Lu} and $f_{\eta_c} = (335 \pm 75)$ MeV \cite{CLEO}, we obtain:
\begin{eqnarray} \nonumber
f_{\psi(2S)} &=& 361.7 \pm 22.5 \; (322.7 \pm 42.7) \ \ \  \text{MeV}, \nonumber \\
f_{\chi_{c1}(1P)} &=& 295.24 \pm 27.48 \; (275.7 \pm 58.5) \ \ \   \text{MeV}, \\  \nonumber
f_{\eta_c(2S)} &=& 203.03 \pm 102.12 \ \ \  \text{MeV}, \\ \nonumber
f_{\chi_{c0}(1P)} &=& 127.4 \pm 38.1 \ \ \  \text{MeV}.  \nonumber
\end{eqnarray}

  From these values we obtain $f_{\eta_c}/f_{\eta_c(2S)}= 1.65 \pm 0.9$ and $f_{J/\psi}/f_{\psi(2S)} = 1.15 \pm 0.07 (1.29 \pm 0.17)$ while in the Ref. \cite{Colangelo-Fazio-Wang} is obtained $f_{\eta_c}/f_{\eta_c(2S)} =f_{J/\psi}/f_{\psi(2S)} = 1.41$. \\

\noindent
\textbf{5.2} \textit{$B^+ \to K^+(\pi^+),J/\psi$ decays}: An important  test to naive factorization is given by
\begin{equation}\nonumber
\frac{\Gamma(B^+ \to K^+,J/\psi)}{\Gamma(B^+ \to \pi^+,J/\psi)} = (18.31 \pm 1.51)  \left| \frac{F_1^{B \to K}(m^2_{J/\psi})}{F_1^{B \to \pi}(m^2_{J/\psi})}  \right|^2 =33.21 \pm 5.14,
\end{equation}

where the errors  come from the numerical values of the CKM  and  form factors (which are evaluated in the CLFA model \cite{CLF}).  The experimental value  of this ratio  is $20.7 \pm 1.8$ \cite{pdg}. This sizable difference means that these exclusive channels have large nonfactorizable contributions \cite{Jung-Mannel}. Some authors have explored the possibility of new physics in these decays \cite{Fleischer-Mannel}.\\

\noindent
\textbf{5.3} \textit{$\overline{B_s^0} \to D_s^+(K^+), K^-(D_s^-)$ decays:} The ratio between the branching ratios of  $\overline{B_s^0} \to D_s^+, K^-$ (mediated by the $b \to c \overline{u}s$ transition) and $\overline{B_s^0} \to K^+,D_s^-$ decays (mediated by the $b \to u \overline{c}s$ transition), which are color favored, is

\begin{equation}\nonumber
\mathcal{R}= \frac{\mathcal{B}(\overline{B_s^0} \to D_s^+, K^-)}{\mathcal{B}(\overline{B_s^0} \to K^+,D_s^-)} = (3.94) \left(  \frac{f_{K^-}}{f_{D_s^-}}   \right)^2  \left| \frac{F_0^{B_s \to D_s}(m^2_{K^-})}{F_0^{B_s \to K}(m^2_{D_s^-})} \right|^2.
\end{equation}

This ratio is  sensitive to the value of the decay constant $f_{D_s^-}$. Evaluating the form factors in the CLFA model  \cite{CLF}, we obtain $\mathcal{R}=9.82 \pm 1.27 \; (11.34 \pm 1.43)$  with $f_{D_s^-} = 259 \pm 7$ \cite{CLEOfDs} $\;(241 \pm 3$  \cite{HPQCD}) MeV. The  sources of the uncertainty come from the CKM factors, the decay constants and the form factors.  The dominant error comes from the value of $V_{ub}$. From  the experimental value $\mathcal{B}(\overline{B_s^0} \to D_s^{\pm}K^{\mp}) = (3.0 \pm 0.7) \times 10^{-4}$ \cite{pdg} it is obtained  $\mathcal{R}=1$ while we compute  $\mathcal{R} \approx 10$.    Thus, with improved measurements, this ratio is a good test to the  numerical inputs for $V_{ub}$ and   $f_{D_s^-}$.
\\

\noindent
\textbf{5.4} \textit{$H \to P'P$ and $P \to l\nu_{l}$ decays:} Let us  compare the two-body nonleptonic  $H(\overline{q}q_1) \to P'(\overline{q}q_2), P(\overline{q_3}q_4)$ and the leptonic $P(\overline{q_3}q_4) \to l\nu_{l}$ decays. It is well known that the decay rate of $P(\overline{q_3}q_4) \to l\nu_{l}$ is

\begin{equation}\nonumber
\Gamma(P \to l\nu_{l}) =  \frac{G_F^{2}|V_{q_3q_4}|^{2}f_P^{2}m_Pm_l^{2}}{8\pi}\left(  1 - \frac{m_l^{2}}{m_P^{2}}  \right)^{2}.
\end{equation}

The ratio between $\Gamma(H \to P', P)$ and  the last expression is given by

\begin{equation}
\frac{\Gamma(H \to P', P)}{\Gamma(P \to l\nu_{l})} = \frac{|V_{q_1q_2}|^{2}a_1^{2}}{4}\frac{(m^{2}_H - m^{2}_{P'})^{2}\lambda^{\frac{1}{2}}(m^{2}_H,m^{2}_{P'},m^{2}_{P})}{m^{3}_H m_l^{2}m_P(1 - \frac{m_l^{2}}{m^{2}_{P}})^{2}} \left|  F_0^{H \to P'}(m^{2}_{P})   \right|^{2}.
\end{equation}

This quotient is independent of the decay constant $f_P$, and could be used as a test for  the form factor  $F_0^{H \to P'}(m^{2}_{P})$. For some exclusive channels, we obtain

\begin{eqnarray}\nonumber
 \left|  F_0^{B^- \to D^0}(m^{2}_{D_s^-}) \right|^{2} &=& 0.301 \pm 0.037 \; (0.293 \pm 0.053),  \\ \nonumber
 \left|  F_0^{B_s^0 \to K^+}(m^{2}_{D_s^-}) \right|^{2} &=& 0.765 \pm 0.216 \;(0.681 \pm 0.197), \nonumber
\end{eqnarray}

with  $l=\tau^- \; (\mu^-)$. The error comes basically from the experimental value of the branching ratios.  We can see that the value of  $|  F_0^{B \to D}(m^{2}_{D_s})|^{2}$ is approximately equal when the lepton $l$ is $\tau$ or $\mu$. The situation for $|  F_0^{B_s \to K}(m^{2}_{D_s})|^{2}$ is different because the value of $V_{ub}$ also is a source of uncertainty. On the other hand, the value of $F_0^{B_s \to K}$ in $q^2=0$  depends strongly on   phenomenological models, ranges from 0.23 to 0.31 \cite{Lu-Yuan-Wei}. Thus, the improvement of   these experimental ratios in future experiments, as LHCb, will be a test of the respective form factors.
\\


\noindent
\textbf{5.5} \textit{$H \rightarrow P_1,P_2(V')$  decays:} Another important ratio is given by the   decay widths of $H \rightarrow P_1,P_2$ and $H
\rightarrow P_1,V'$, where $P_2$ and $V'$ have the same quark content with  $P_1 = P, \; S, \; P(2S)$, $P_2 = P, \; P(2S)$ and $V' = V, \; A(^3P_1), \; V(2S)$. Using the expressions given in  Table III and   monopolar form factors with the fact that $F_{0}^{H \rightarrow P_1}(0)=F_{1}^{H \rightarrow P_1}(0)$ \cite{wsb}, we obtain:

\begin{equation}\label{11}
\frac{\Gamma(H \rightarrow P_1,P_2)}{\Gamma(H \rightarrow P_1,V')} =
\left(\frac{f_{P_2}}{f_{V'}} \right)^{2} \left[
\frac{1-m_{V'}^{2}/m_{1^{-}}^{2}}{1-m_{P_2}^{2}/m_{0^{+}}^{2}}\right]^{2}
\frac{\left[\lambda(m_{H}^{2},m_{P_1}^{2},m_{P_2}^{2})\right]^{1/2}}{\left[\lambda(m_{H}^{2},m_{P_1}^{2},m_{V'}^{2})\right]^{3/2}}
(m_{H}^{2}-m_{P_1}^{2})^{2}.
\end{equation}

This ratio provides information on the quotient $f_{P_2}/f_{V'}$. As an example, we obtain $(f_{\pi^{+}}/f_{\rho^{+}})=0.631 \pm 0.045$ using the    $B^{0} \rightarrow D^{-},\pi^{+}$ and $B^{0} \rightarrow D^{-},\rho^{+}$ decays which  branching ratios are $(2.68 \pm 0.13)\times 10^{-3}$ and $(7.6 \pm 1.3)\times10^{-3}$, respectively \cite{pdg}.  The main uncertainty arises from these experimental values. On the other hand,  taking $f_{\pi^{+}}=(130.7 \pm 0.4)$ MeV and $f_{\rho^{+}}= (216 \pm 2)$ MeV  \cite{CLF} it is obtained $(f_{\pi^{+}}/f_{\rho^{+}})=0.605 \pm 0.006$. So, in this case factorization assumption gives a good approximation to the value of this quotient.   \\

\noindent
\textbf{5.6} \textit{$H \to P', V_{1(2)}$ decays:} In order to obtain  $f_{V_1}/f_{V_2}$, we can  consider the  ratio between the decay rates of $H \to P', V_1(q_i\overline{q_j})$ and $H \to P',V_2(q_i\overline{q_j})$, where $P' = P, \;  S, \; P(2S)$ and $V_{1,2} = V, \; A(^3P_1), \; V(2S)$:

\begin{equation}
\frac{\Gamma(H \rightarrow P',V_1)}{\Gamma(H \rightarrow P',V_2)} =
\left(\frac{f_{V_1}}{f_{V_2}} \right)^{2} \left|    \frac{F_1^{H \to P'}(m^2_{V_1})}{F_1^{H \to P'}(m^2_{V_2})}           \right|^2 \left[\frac{\lambda(m_{H}^{2},m_{P'}^{2},m_{V_1}^{2})}{\lambda(m_{H}^{2},m_{P'}^{2},m_{V_2}^{2})}\right]^{3/2}.
\end{equation}

Let us choose, as an application,   the $B \rightarrow P,V$ and $B
\rightarrow P,A$ processes.  From the expressions in  Table III and using  monopolar form factors \cite{wsb}   we obtain:

\begin{equation}
\frac{\Gamma(B \rightarrow P,V)}{\Gamma(B \rightarrow P,A)} =
\left(\frac{f_{V}}{f_{A}} \right)^{2} \left[
\frac{1-m_{A}^{2}/m_{1^{-}}^{2}}{1-m_{V}^{2}/m_{1^{-}}^{2}}\right]^{2}
\left[\frac{\lambda(m_{B}^{2},m_{P}^{2},m_{V}^{2})}{\lambda(m_{B}^{2},m_{P}^{2},m_{A}^{2})}\right]^{3/2}.
\end{equation}

Taking the   $B^{0} \rightarrow D^{-},\rho^{+}$ and $B^{0} \rightarrow D^{-},a_1^{+}$ decays  we get $(f_{\rho}/f_{a_1})=1.06 \pm 0.31$. The dominant error comes from  the experimental value  $\mathcal{B}(B^{0} \rightarrow D^{-}a_1^{+})=(6.0 \pm 3.3)\times 10^{-3}$.  With $f_{\rho}=(216 \pm 2)$ MeV \cite{CLF} it is obtained $f_{a_1}=(0.203 \pm 0.059)$ GeV. This value is smaller than the one reported in the literature. For example, in the Ref. \cite{cheng-chiang}, $f_{a_1}=0.238 \pm 0.010$ GeV while the Ref. \cite{neubert} gives $f_{a_1}=\text{0.229}$ GeV (extracted from the  $\tau^{-} \rightarrow M^{-}\nu_{\tau}$ decay) and $f_{a_1}=\text{0.256}$ GeV (from the $\overline{B^{0}} \rightarrow
D^{*+},a_{1}^{-}$  and $\overline{B^{0}} \rightarrow
D^{*+},\rho^{-}$ decays). On the other hand, Ref. \cite{nardulli} obtained $f_{a_1}=0.215 \; (0.223)$ GeV for $\theta = 32^{\circ}\; (58^{\circ})$, where $\theta$ is the mixing angle between the $K_{1A}$ and $K_{1B}$ mesons.  As the error in $\mathcal{B}(B^{0} \rightarrow D^{-}a_1^{+})$ is too big,  it is important to get a more precise estimation  of this branching in future experiments  in order to test hypothesis factorization with these exclusive decays. \\

It is also possible to obtain the quotient  $(f_{\rho}/f_{a_1})$ from  $\mathcal{B}(\overline{B_s^0} \to D_s^+, \rho^-)/\mathcal{B}(\overline{B_s^0} \to D_s^+, a_1^{-})$ and $\mathcal{B}(B_c^{-} \to \eta_c, \rho^-)/\mathcal{B}(B_c^{-} \to \eta_c, a_1^{-})$. At  present, there are not  experimental values of these branchings. So, in the future these decays will be a test of naive factorization by means of the ratio  $(f_{\rho}/f_{a_1})$. \\


\noindent
\textbf{5.7} \textit{$H \rightarrow V_1,V_{2(3)}$ decays:} Another important ratio in order to compute the quotient $f_{V_1}/f_{V_2}$ comes from the $H \rightarrow V_1,V_2(q_i\overline{q_j})$ and $H
\rightarrow V_1,V_3(q_i\overline{q_j})$ processes, where $V_1 = V, \; A(^1P_1), \;  A(^3P_1), \;  V(2S)$ and $V_{2,3} = V,  \; A(^3P_1), \; V(2S)$.  As an example, we consider the $B \to V, V'$ and $B \to V, A$ decays. From expressions displayed in  Table III we obtain:

\begin{equation}\label{17}
\frac{\Gamma(B \rightarrow V,V')}{\Gamma(B \rightarrow V,A)} =
\left(\frac{f_{V'}}{f_{A}} \right)^{2} \frac{\mathcal{G}(m_{V'}^{2})}{\mathcal{G}(m_{A}^{2})}.
\end{equation}

Taking the  $B^{0} \rightarrow D^{*-},\rho^{+}$ and $B^{0} \rightarrow D^{*-},a_1^{+}$ decays and evaluating $\mathcal{G}$ with  appropriate monopolar form factors \cite{wsb}, we get  $(f_{\rho}/f_{a_1})=0.81 \pm 0.07$, where the source of uncertainty are the form factors. This value agrees with the one reported in Ref. \cite{neubert}, although is smaller than the value obtained in previous subsection.  \\

$f_{\rho}/f_{a_1}$ can also be obtained from $\Gamma(\overline{B_s^0} \to D_s^{*+}, \rho^-)/\Gamma(\overline{B_s^0} \to D_s^{*+}, a_1^-)$ and $\Gamma(B_c^{-} \to J/\psi, \rho^-)/\Gamma(B_c^{-} \to J/\psi, a_1^{-})$. At  present, there is not experimental information of these decays in order to test the factorization hypothesis.  \\

\noindent
\textbf{5.8} \textit{$B \rightarrow M_{1},M_{2}$ and $B \rightarrow M_{1}l\nu_{l}$ decays:} It is well known that the ratio $\emph{R} = \Gamma(B \rightarrow M_{1},M_{2})/[d\Gamma(B \rightarrow M_{1}l\nu_{l})/dt|_{t=m^{2}_{M_2}}]$ provides a method to test factorization hypothesis and may be used to determine some unknown decay constants \cite{neubert,ratio}. Also, it is possible combining exclusive semileptonic and hadronic $B$ decays to measure CKM matrix elements (see for example Ref. \cite{pseudoscalar}). If  $M_1$ is  any of the eight mesons showed in  Table I, $M_2(q_i\overline{q_j})$ is a $J=1$ meson and $m_l \approx 0$, we obtain

\begin{equation}\label{18}
R_{V'} =  \frac{\Gamma(H \to M,V')}{d\Gamma(H \to
Ml\nu_{l})/dt|_{t=m^2_{V'}}} = \frac{\xi^{(V')}}{\zeta} =
6\pi^{2}|V_{ij}|^{2}(a_{1}^H)^{2}f_{V'}^{2},
\end{equation}
where $V' = V, \; A(^3P_1), \; V(2S)$. Thus, $R_{V'}$, which is model-independent,  is a clean and direct test of factorization hypothesis. On the other hand, assuming the validity of the factorization with a fixed value for $a_1$, it provides an alternative use: it may be used for  determination of unknown decay constants.   For example,  $f_{\rho}$ can be obtained from
\begin{equation}\nonumber
R_{\rho^-} \equiv \frac{\Gamma(B^- \to D^0,\rho^-)}{d\Gamma(B^- \to
D^0l\nu_{l})/dt|_{t=m_{\rho}^{2}}} = \frac{\Gamma(B_s^0 \rightarrow D_s^+,\rho^-)}{d\Gamma(B_s^0 \rightarrow
D_s^+l\nu_{l})/dt|_{t=m_{\rho}^{2}}} = \frac{\Gamma(B_c^- \rightarrow \eta_c,\rho^-)}{d\Gamma(B_c^- \rightarrow
\eta_cl\nu_{l})/dt|_{t=m_{\rho}^{2}}},
\end{equation}
where $R_{\rho^-} = 6\pi^{2}|V_{ud}|^{2}(a_{1}^H)^{2}f_{\rho^-}^{2}$.\\

We also can use the equation (\ref{18}) in order to obtain ratios between decay constants of $J=1$ mesons:

\begin{equation}
\frac{{R_{V_1'}}}{R_{V_2'}} = \left( \frac{f_{V_1'}}{f_{V_2'}} \right)^2 , \ \ \ \ V_{1,2}' = V, \; A(^3P_1), \; V(2S).
\end{equation}
\\

\noindent
\textbf{5.9} \textit{$\overline{B_{(s)}^0} \to D_{(s)}^+, \pi^-(K^-)$ decays:} Taking  pairs of decays that are U-spin\footnote{The U-spin symmetry is a SU(2) subgroup of the SU(3) flavor symmetry group, in which quarks $d$ and $s$ form a doublet \cite{Jung-Mannel, Uspin}.}  partners, we get

\begin{equation}\nonumber
\mathcal{R}_{\pi/K}= \frac{\mathcal{B}(\overline{B_s^0} \to D_s^+, \pi^-)}{\mathcal{B}(\overline{B^0} \to D^+, K^-)} = (12.45) \left| \frac{F_{0}^{B_s \to D_s}(m^{2}_{\pi})}{F_{0}^{B \to D}(m^{2}_{K})}  \right|^{2} = 13.07 \pm 0.32
\end{equation}
and  $\mathcal{R}_{K/\pi} = 0.082 \pm 0.002$,  evaluating  the form factors  in the CLFA model \cite{CLF}. The dominant source of error comes from these  form factors. In the first (second) case, the ratio between the  experimental values of the branching ratios    \cite{pdg} is $16.0 \pm 5.4 \; (0.112 \pm 0.027)$. In both cases, the experimental ratio is bigger than the theoretical one.  Therefore, with improved measurements at future experiments as LHCb, these ratios will be  a good test of the breaking of U-spin symmetry  through the ratio of the form factors. On the other hand, they  provide an alternative strategy in order to determine $f_K/f{\pi}$ and compare with other methods (see for example  Ref. \cite{pi-K}).

\section{Summary}

We computed several useful ratios between decay widths of two-body nonleptonic and semileptonic $B$ and $B_s$ decays, which with improved measurements in forthcoming experiments as LHCb, could be test of factorization approach by means of quotients between form factors or  decay constants. The ratios with $B$ decays considering  charmonium states and  light mesons in final state (see subsection 5.1)  could be  the more likely scenario to test the factorization scheme. It is important to mention that divergences  from the results obtained  assuming  the current approximations do not imply a failure of  the QCD itself or the factorization approach alone.  It would be required a more exhaustive and comprehensive analysis  for  getting more  conclusions on these and possible new physics effects in these decays.   We also presented a summary of the expressions for $\Gamma(H \rightarrow M_1,M_2)$ and $d\Gamma(H \rightarrow M_1l\nu)/dt$, at tree level, including eight types of mesons in final state: $M_{1,2}$ can be a ground state meson ($l=0$), or an orbitally excited meson ($l=1$) or a radially excited meson ($n=2$), assuming  factorization hypothesis and using  the parametrizations of $\langle M|J_{\mu}|B\rangle$ given in  the WSB and the ISGW quark models. The  form factors were evaluated  in the WSB and CLFA quark models. We classified in three groups the  $H \to M_{1,2}$ transitions and  explained some aspects related with  the dynamics of these processes.

\begin{center}
\textbf{Acknowledgements}
\end{center}

This work has been partly supported by Comit\'e Central de Investigaciones of University of Tolima. J. H. Mu\~noz also thanks CNPq (Brazil) for financial support.

\appendix

\section*{Appendix}

In this appendix, we briefly mention  the quark models and their form factors that are  used in this work.\\
 
\textit{1. The ISGW model \cite{isgw}:} it is a hybrid model that combines  a nonrelativistic quark potential model with a phenomenological ansatz. It is consistent  with heavy quark symmetry at maximum recoil $t_m$. Their  form factors are modeled by a gaussian and normalized at $q^2_{max}$.  All the form factors in this model are in function of 
\begin{equation}
F_n^{H \to M}(q^2) = \left( \frac{\tilde{m}_M}{\tilde{m}_H}  \right)^{\frac{1}{2}}  \left(  \frac{\beta_H \beta_M}{\beta^2_{HM}}  \right)^{\frac{n}{2}} e^{- \Lambda(t_m - q^2)},
\end{equation}
where $\Lambda = m_d^2/(4 \kappa^2 \tilde{m}_H \tilde{m}_M \beta^2_{HM})$.  $\tilde{m}_{H(M)}$ is the mock mass of the $H(M)$ meson, $\beta$ is a variational parameter and $\kappa = 0.7$ is a relativistic compensation factor of the model. The appendix B of the Ref. \cite{isgw} has all the required inputs for evaluating the form factors for the $H \to M(J =0, 1, 2)$ transition.\\

2.  \textit{The WSB  model \cite{wsb}:} It gives the form factors in terms of relativistic bound state wave functions taking the solutions from a relativistic harmonic oscillator potential. The form factors are calculated as wave function overlaps in the infinite momentum frame at $q^2=0$. The mononopolar form factors in this model present a vector meson dominance form of the $q^2$ dependence  and are  given by 
\begin{equation}
F^{H \to M}(q^2) = \frac{F^{H \to M}(0)}{1-q^2/m_{J^P}^{2}}, 
\end{equation}
where $m_{J^P}$ is the mass of the pole. The Ref. \cite{wsb} provides the values of $F_{n}^{H \to M}(0)$ and $m_{J^P}$ for the $H \to M$ transition. We use these form factors in order to compute the numerical values showed in  subsections 5.5, 5.6 and 5.7.\\

We can  obtain  the form factors of the WSB model \cite{wsb} in function of  the form factors of the ISGW model \cite{isgw} comparing  the parametrizations given in both models for the $H \rightarrow P (V)$ transition. Making $\langle P| J_{\mu}|H\rangle_{WSB}$ = $\langle P| J_{\mu}|H\rangle_{ISGW}$ we obtain:
\begin{eqnarray}
F_{0}(t) &=& \frac{t}{(m_{H}^{2}-m_{P}^{2})} \;
f_{-}(t) \; + f_{+}(t), \\
F_{1}(t) &=& f_{+}(t),
\end{eqnarray}
and from $\langle V| J_{\mu}|H\rangle_{WSB}$ = $\langle V| J_{\mu}|H\rangle_{ISGW}$ it is obtained:
\begin{eqnarray}
A_{0}(t) &=& \frac{i}{2m_{V}} \left[f(t)+ ta_{-}(t)+
(m_{H}^{2}-m_{V}^{2})a_{+}(t) \right],\\
A_{1}(t) &=& \frac{if(t)}{(m_{H} + m_{V})}, \\
A_{2}(t) &=& -i(m_{H}+ m_{V}) \;a_{+}(t), \\
V(t) &=& -i(m_{H}+m_{V})\; g(t).
\end{eqnarray}
Using these relations it is straightforward to get $d\Gamma(H \rightarrow P(V)l\nu)/dt$ or $\Gamma(H \rightarrow P(V),M)$ with the parametrization of the WSB model from  respective expressions in the ISGW model, and viceversa.\\

3.  \textit{The CLFA  model \cite{CLF}:} The relativistic light-front quark model gives a fully treatment of quark spin and the center-of-mass motion of the hadron. In a covariant approach of this model the decay constants and the form factors are calculated by means of Feynman momentum loop integrals which are manifestly covariant \cite{CLF}. The form factors in the spacelike region are given by the three-parameter form

\begin{equation}
F^{H \to M}(q^2) = \frac{F^{H \to M}(0)}{1 - a(q^2/m^2_H) + b (q^2/m^2_H)^2}.
\end{equation} 
We have taken from the  Ref. \cite{CLF} the values of $F^{H \to M}(0)$, $a$ and $b$ for obtaining the numerical values presented in subsections 5.1, 5.2, 5.3 and 5.9.

\end{document}